\documentclass[12pt,preprint]{aastex}

\shorttitle{Two Distinct GRB Components}

\shortauthors{J. Hakkila and T. W. Giblin}

\begin{document}




\title{Quiescent Burst Evidence for \\ Two Distinct GRB Emission Components}







\author{Jon Hakkila and Timothy W. Giblin}

\affil{Department of Physics and Astronomy, College of Charleston,
    Charleston, SC 29424-0001}






\begin{abstract}

We have identified two quiescent GRBs (bursts having two or more 
widely-separated distinct emission episodes) in which the post-quiescent 
emission exhibits distinctly different characteristics than the pre-quiescent 
emission. In these two cases (BATSE GRBs 960530 and 980125), the second 
emission episode has a longer lag, a smoother morphology, and softer 
spectral evolution than the first episode. Although the pre-quiescent 
emission satisfies the standard internal shock paradigm, we demonstrate 
that the post-quiescent emission is more consistent with external shocks. 
We infer that some observed soft, faint, long-lag GRBs are external shocks 
in which the internal shock signature is not observed. We further note 
that the peak luminosity ratio between quiescent episodes is not in 
agreement with the ratio predicated by the lag vs. peak luminosity 
relationship. We briefly discuss these observations in terms of current 
collapsar jet models.

\end{abstract}














\section{Introduction}

There is strong evidence that the majority of gamma-ray burst (GRB) pulses 
result from internal shocks in relativistic winds 
\citep{sar97a,kss97,dm98,rrf00,np02}. However, a number of bursts exhibit 
a soft component indicative of external shocks that could be interpreted 
as onset of afterglow \citep{gib99,bur99,gib02}. This emission begins 
preferentially towards the end of the burst or even after the GRB has 
ended (as suggested by the BeppoSAX WFC and NFI observations of serveral 
x-ray afterglows \citep{cos00}), but can also appear at a time early enough 
to overlap the short-timescale emission (as observed in GRB 980923 
\citep{gib99}). In addition, extended soft $\gamma-$ray emission has been 
observed by co-adding the fluxes of many BATSE bursts; this might also be 
an indicator of the same external shock phenomenon \citep{con02}.

``Quiescent'' GRBs may provide new clues to the origins and physics of 
GRB shocks. Quiescent bursts are GRBs that release their gamma-ray energy 
in more than one distinct emission episode \citep{rm01,rmr01}; e.g. they 
have at least one extended period during which emission is absent. A 
rigorous definition of quiescent time can be problematic due to the 
diverse nature or complexity of GRB temporal structures.

\citet{rm01} introduced a quiescent GRB definition for a sample of bright 
BATSE bursts that used a sliding temporal window with a width of 
$0.05 \times {\rm T90}$ on summed, background-subtracted, BATSE 4-channel 
64-ms data. These authors chose GRBs which had at least one time period 
for which the count rate dropped below $2\sigma$ of the summed background 
rate. This criterion was applied to only the brightest, longest bursts in 
the BATSE 4B Catalog \citep{pac99}.

It is possible that this definition is too permissive, since it might allow 
for inclusion of some GRBs that do not have distinct emission episodes. For 
example, some bright bursts that do not satisfy this criterion (because the 
first episode's decay and the second episode's rise overlap) might be 
classified as quiescent if they are observed at lower signal-to-noise, as 
might some bursts with faint, long, underlying pulses. Because the existing 
quiescent definition might be too permissive, we have tentatively further 
required quiescent GRBs to be those in which the emission drops to the 
background for an extended period of time that is greater than or equal 
to the duration of the previous emission episode. This ensures that the 
quiescent time is greater than the interpulse duration.

%Long decay tails can also affect the quiescent 

%definition through the T90 measurement. This is 

%because T90 is relatively insensitive to faint 

%extended emission, so that bursts with long tails 

%can have significantly shorter T90 values than 

%might be obtained using some other duration measure. 

%There is thus a potential classification bias 

%because the quiescent GRB definition depends on T90. 

%The definition is also less successful at 

%identifying faint quiescent GRBs than bright ones. 

%This is because inter-episodic emission is less 

%likely to be observed in low signal-to-noise data. 

%This can cause faint non-quiescent GRBs to take on 

%quiescent characteristics. Sampling techniques and 

%instrumental characteristics can also play a part 

%in quiescent GRB classification. For example, long 

%integration times increase signal-to-noise ratios, 

%so detection of inter-episodic emission is more 

%likey using a detector with long integration times. 

%Additionally, the definition is dependent upon the 

%choice of energy channels summed as well as upon 

%the spectral characteristics of the emission episodes. 

In their study of very bright quiescent GRBs, \citet{rm01} operated under 
the assumption that the post-quiescent episodes are due to internal shocks. 
We offer the hypothesis that quiescent GRBs potentially provide a laboratory 
in which {\it external} shock signatures indicative of the onset of afterglow 
may be isolated from the rest of the burst. We have thus far found two GRBS 
that support this hypothesis.

\section{Analysis}

GRB 960530 and GRB 980125 are bursts from the BATSE 5B catalog that deviate 
significantly from what is considered to be standard GRB behavior. Neither 
burst was observed by Bepposax, and the RXTE PCA was unable to localize GRB 
980125.

\begin{figure}
%\plotone{5478_hist.eps}
\plotone{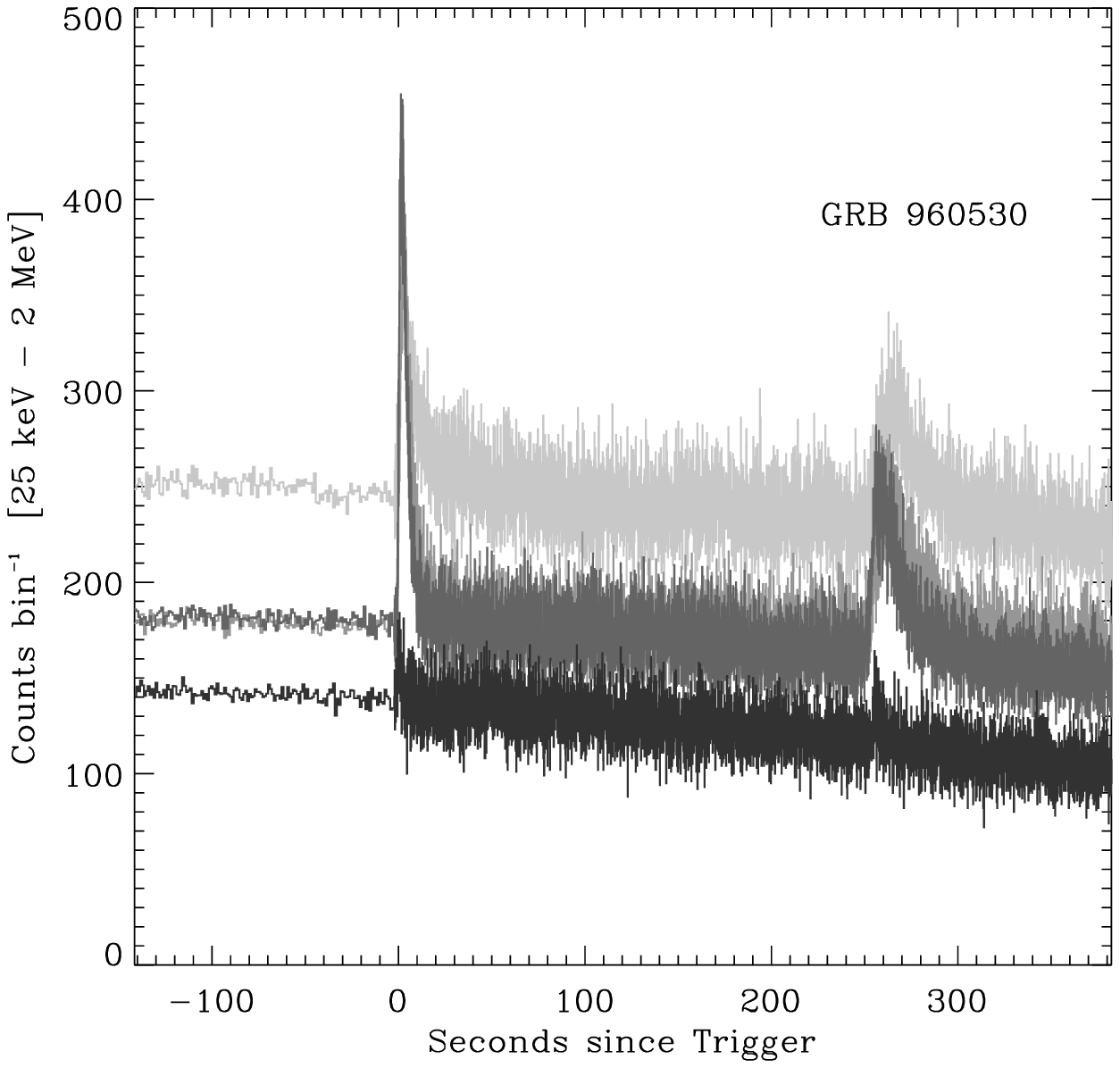}
\caption{Four-channel time history of GRB 960530 (BATSE trigger 5478). \label{fig1}}
\end{figure}

\begin{figure}
%\plotone{6581_hist.eps}
\plotone{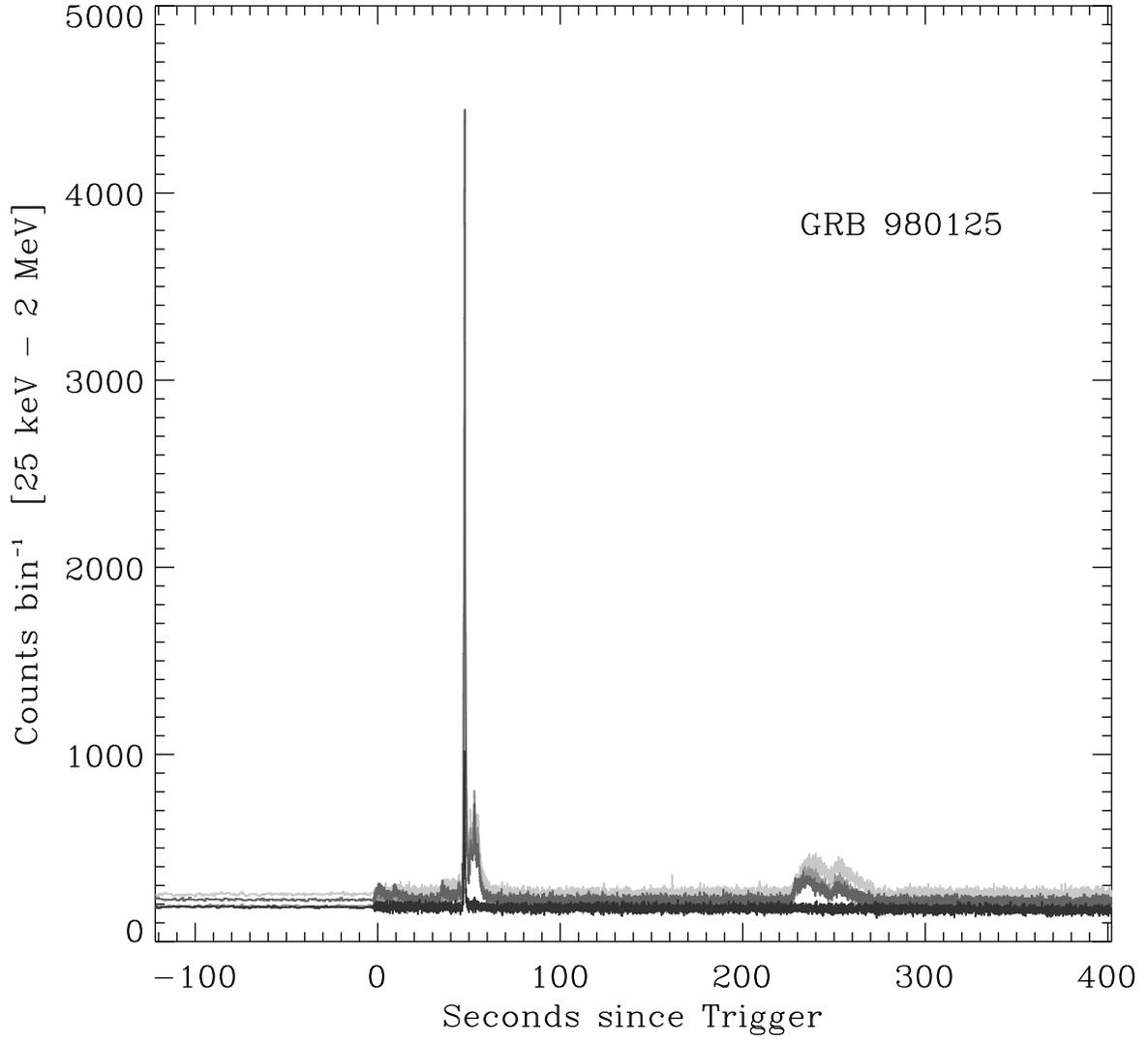}
\caption{Four-channel time history of GRB 980125 (BATSE trigger 6581). \label{fig2}}
\end{figure}

The BATSE 4-channel time histories of GRB 960530 and GRB 980125 are 
illustrated in figure~\ref{fig1} and figure~\ref{fig2}. We have measured 
quiescent and emission episodic durations for these GRBs via the summed 
four-channel sliding temporal window technique of \cite{rm01}. As shown 
below, the localization measurements verify that the multiple emission 
episodes are indeed associated with the same sources for both GRB 960530 
and GRB 980125.

GRB 960530 (BATSE trigger 5478) has two distinct emission episodes separated 
by 233 s. Episode (a) is a single-pulse FRED (Fast Rise Exponential Decay) of 
duration 23 s. Episode (b) is a smooth single-pulse episode of 37 s. The BATSE 
locations of these episodes are consistent with the same source; episode (a) 
is located at $\rm{RA} = 143.0^\circ$ and $\rm{Dec} = -53.8^\circ$ with error 
$\pm 1.4^\circ$, while episode (b) is located at $\rm{RA} = 143.0^\circ$ and 
$\rm{Dec} = -53.0^\circ$ with error $\pm 1.9^\circ$.

GRB 980125 (BATSE trigger 6581) has two distinct emission episodes with a 
faint preburst episode. The preburst episode is faint and lasts 13 s. It 
is followed by a short quiescent period of 18 s. Although the preburst 
episode satisfies our rigorous definition of a distinct emission episode, 
we have chosen not to include it in our analysis because it is too faint 
for its properties to be accurately measured. Complex emission episode (a) 
follows and lasts 31 s. Episode (b) is a smooth episode with two distinct 
pulses and a duration of roughly 45 s, it is preceded by a quiescent period 
of 161 s. The first of these two distinct pulses is a smooth broad pulse 
similar in temporal structure to GRB 960530's emission episode (b), while 
the second pulse (which overlaps the first) is more indicative of a standard 
FRED. The BATSE locations of these episodes are also consistent with the same 
source; the faint preburst pulse is located at $\rm{RA} = 344.2^\circ$ and 
${\rm Dec} = 41.0^\circ$ with an error of $\pm 3.8^\circ$, emission episode 
(a) is located at ${\rm RA} = 350.6^\circ$ and $\rm{Dec} = 35.1^\circ$ with 
an error of $\pm 0.5^\circ$, and emission episode (b) is located at 
$\rm{RA} = 354.6^\circ$ and $\rm{Dec} = 34.8^\circ$ with error $\pm 1.4^\circ$.

We demonstrate that the smooth quiescent pulses have other peculiar properties 
in addition to their peculiar temporal properties. To study these quiescent 
GRBs in more detail, we apply three techniques: (1) spectral lags, (2) 
Color-Color Diagrams (CCDs), and (3) the Internal Luminosity Function (ILF).

\subsection{Spectral Lags}

Spectral lags are energy-dependent delays in the GRB temporal structure. In 
general, soft pulses lag behind hard pulses. Pulses also appear to broaden 
at softer energies as $E^{-0.4}$ \citep{nor96}. Spectral lags apparently 
correlate with burst parameters such as beaming angles and peak luminosities 
(as measured on the 256 ms timescale) \citep{nor02}.

Spectral lags are defined as the maximum value of the Cross-Correlation 
Function (CCF) of two energy channels \citep{ban97}. Poisson background 
variations in the time histories complicate the identification of the 
CCF peak, so prior studies have modeled the CCF with a fitting function 
to identify the lag. CCFs are typically asymmetric because GRBs generally 
have longer decay times than rise times. A cubic fitting function has been 
found to be a more accurate characterization of CCF lags \citep{nor00} than 
a quadratic function \citep{wu00}. 

Rather than use a cubic or quadratic fitting function, we have instead 
chosen to use the pulse model of \citet{nor96}. This function is a more 
natural choice than the cubic or quadratic since the CCF describes a 
normalized comparison between episodes whose primary structures are 
pulses; it thus accounts for the CCF time-asymmetry in a natural way. 

Application of the CCF to the different emission components of GRB 960530 
and GRB 980125 leads to intriguing results. The smooth pulse appearing after 
the quiescent period in GRB 960530 has a significantly longer lag (in all 
energy channel combinations) than the first FRED pulse. The lags are shown 
in table~\ref{tbl-1}. The results are similar and are shown in 
table~\ref{tbl-2}) for GRB 980125: the smooth pulse in emission episode (b) 
has a broader CCF and a significantly longer lag than that found during the 
rest of the burst. It should be noted that the lag of emission episode (a) 
is consistent with the lag of the second (FRED) pulse found in emission 
episode (b).

\subsection{The Internal Luminosity Function}

The Internal Luminosity Function (ILF) $\psi(L)$ is the distribution of 
luminosity within a GRB; $\psi(L) \Delta L$ represents the fraction of 
time during which a burst's luminosity lies between $L$ and $L+ \Delta L$ 
\citep{hor97, hak04}. For the purposes of this manuscript, we discuss the 
ILF calculated in the 50 to 300 keV energy range using four-channel 64-ms 
data. The ILF is normalized to the number of Poisson background-subtracted 
time intervals during which emission is observed $2 \sigma$ or more above 
the background and by the requirement that $\Sigma \psi (L) \Delta L =  1$. 
It is best fit with a quasi power-law form for most bursts such that 
\begin{equation}
\psi(L) \propto L^{\alpha} \times 10^{\beta \times [\log L]^2}.
\end{equation}
We refer to $\alpha$ as the power-law index and $\beta$ as the curvature 
index. The variables $\alpha$ and $\beta$ appear to be indicators of burst 
temporal morphology (they are highly-correlated such that GRBs with large 
$\alpha$ also have large $\beta$ while GRBs with small $\alpha$ also have 
small $\beta$); smooth, FRED-like bursts typically have large $\alpha$ 
(roughly $-2 \leq \alpha < 3$) while spiky (narrow-pulsed) bursts typically 
have small $\alpha$ (approximately $-7 \leq \alpha < -2$).

We have measured the ILF for the different components of quiescent GRBs 
960530 and 980125; these are indicated in table~\ref{tbl-1} and 
table~\ref{tbl-2}. We note that the late-appearing, smooth, long-lag 
components of both bursts are fit by noticeably steeper ILF values than 
are the other episodes. The ILF values of these smooth pulses thus indicate 
a different morphology than that normally associated with FREDs. The ILF 
thus supports the hypothesis that the smooth broad pulses are different 
in nature than the pulses found elsewhere in these two quiescent GRBs.

The second (FRED) pulse in emission episode (b) (at time t=255 s) of GRB 
980125 is somewhat peculiar. It is smoother than a typical FRED and has a 
correspondingly steep value of $\alpha$, but it has a short lag consistent 
with the burst's emission episode (a). It is possible that overlap with 
the broad, long-lag pulse has affected its ILF measurement.

\subsection{Color-Color Diagrams}

Color-color diagrams (CCDs) \citep{gib00} can be used to describe GRB 
spectral evolution. Several colors can be created from BATSE four-channel 
data; \cite{gib00} defines the hard color index ${\rm HC} = F3/F2$ where 
F3 is the BATSE channel 3 photon flux in units of photons cm$^{-2}$ s$^{-1}$ 
and F2 is the BATSE channel 2 photon flux. The soft color index SC is defined 
similarly as ${\rm SC} = F2/F1$ where F1 is the BATSE channel 1 photon flux. 
The timescales over which these color indices are measured vary in order to 
maintain a roughly constant signal-to-noise ratio. The BATSE color indices 
in general have large enough signal-to-noise ratios for reliable spectral 
evolution studies. Given the detector response, the observed colors of 
cooling external shocks have been predicted by \cite{gib00}. With this 
information, CCDs can be used to search GRBs for signatures of external 
shocks. Having ascertained that the temporal and temporally-shifted 
spectral structures of the broad pulses are different than those of 
typical GRB pulses, we now explore the spectral evolution of these 
pulses using CCDs. 

GRBs typically undergo hard to soft spectral evolution. Thus, the terminal 
GRB emission tends to be softer than the initial emission. This general 
spectral evolution can be observed in a CCD.

Although the emission appearing towards the end of GRBs tends to be soft, 
not all terminal emission exhibits the colors predicted by the external 
shock model. Theoretical synchrotron cooling models are predicted to 
occupy a fairly narrow range of hard vs. soft colors. Both the fast 
cooling and slow cooling models are predicted to have their color indices 
evolve in the regimes $0.6 \leq SC \leq 1.0$ and $0.7 \leq HC \leq 1.1$ 
\citep{gib00,gib02}.

\begin{figure}
\plottwo{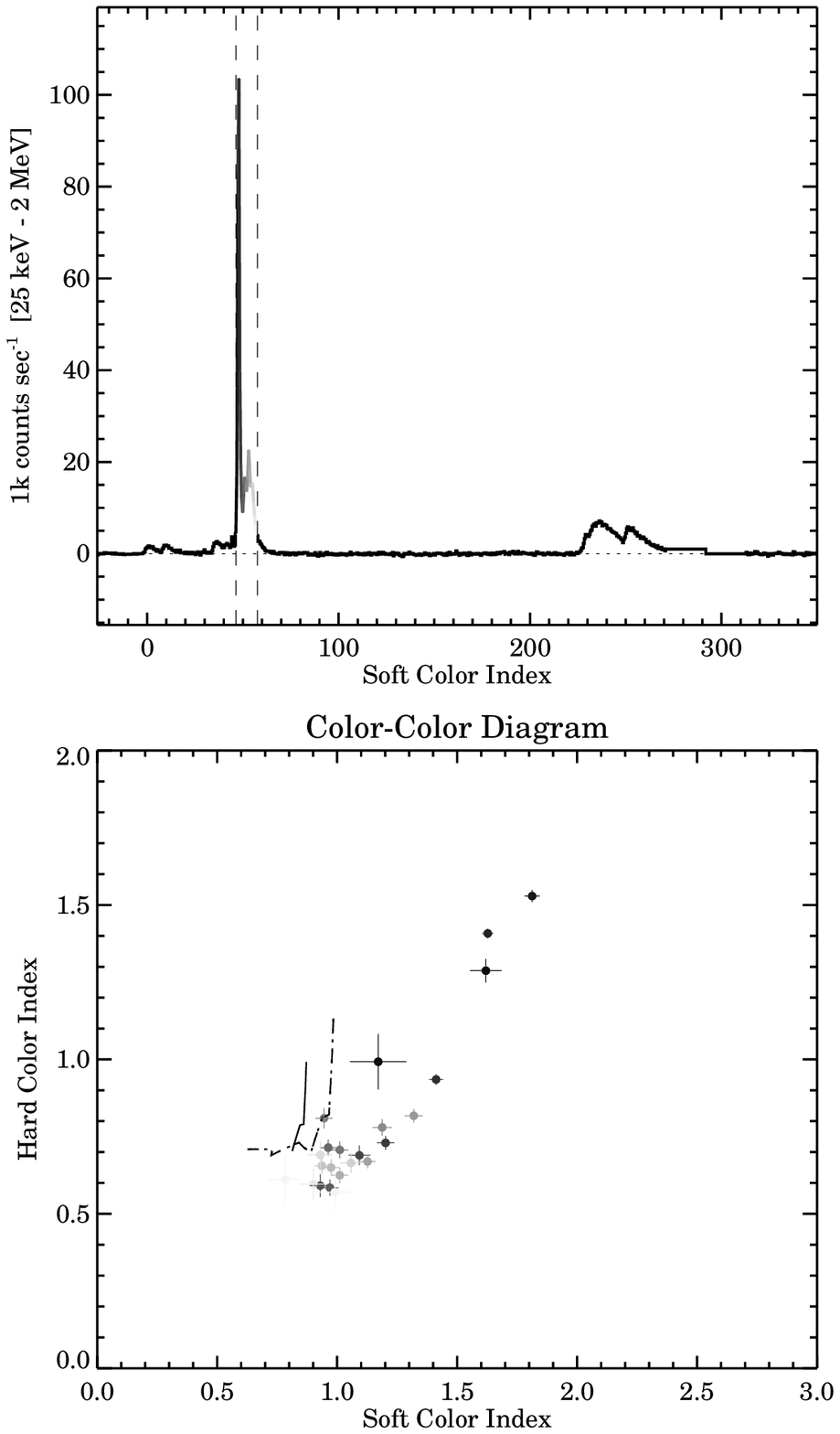}{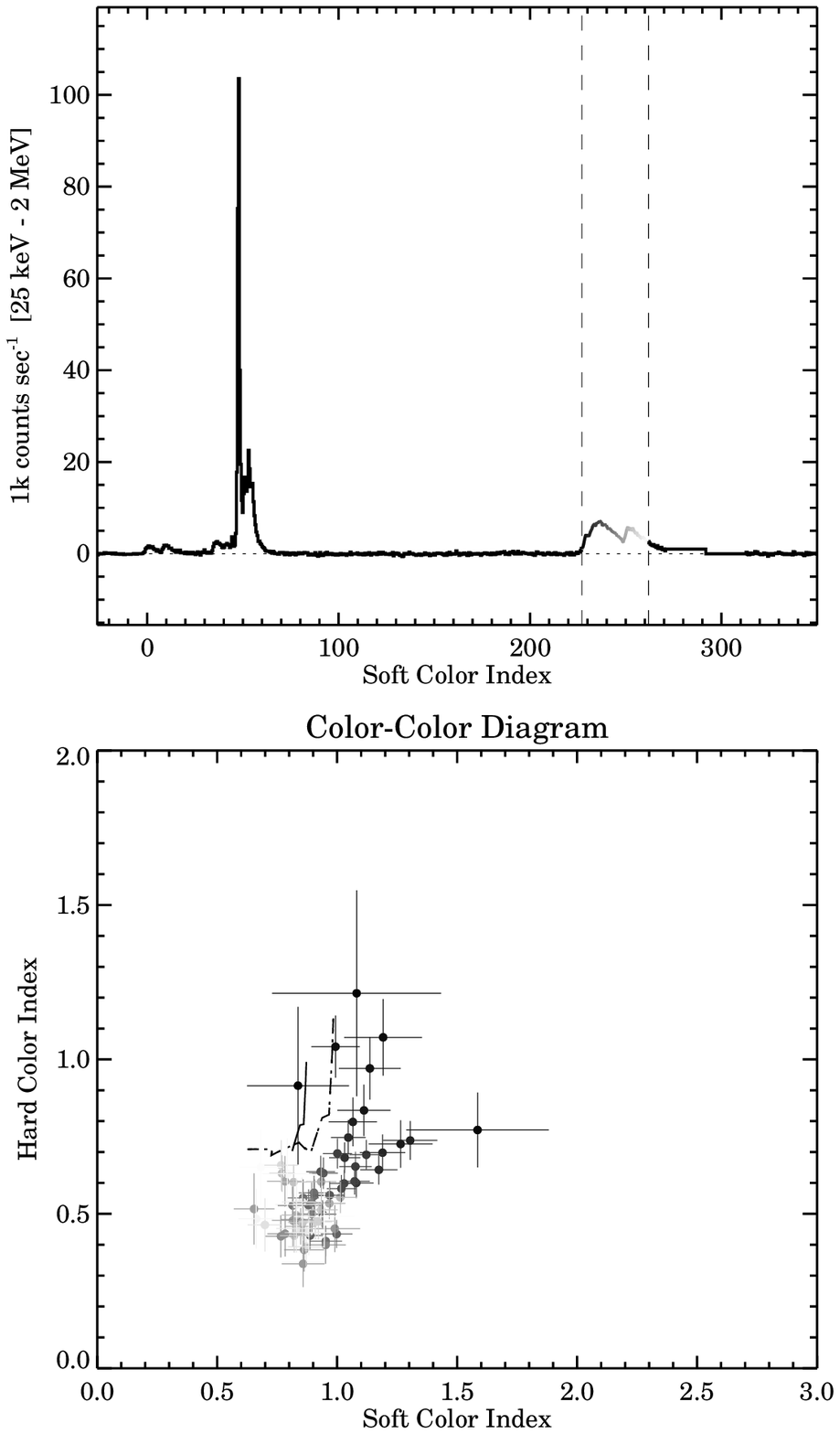}
\caption{Color-Color diagram for the first and second emission episodes of GRB 980125 (BATSE trigger 6581).\label{fig3}}
\end{figure}

GRB 960530 has previously been considered to be an afterglow candidate 
because it exhibits a soft gamma-ray component that was coincident with 
the onset of the afterglow phase \citep{gib02}. The CCD of this soft 
gamma-ray component was additionally in agreement with the theoretical 
prediction for synchrotron cooling. It is difficult to firmly establish 
the connection between the CCD of late emission from GRB 980125 and the 
synchrotron shock model, but figure~\ref{fig3} indicates that this late 
emission is consistent with both fast- (solid line), and slow-cooling 
(dashed line) synchrotron models. Furthermore, the second episode is 
in better agreement with synchrotron afterglow emission than the first 
episode \citep{fre03}.

The concluding time periods of GRB 960530 and GRB 980125 thus both exhibit 
spectral behavior that evolves from hard to soft (such as is typically found 
in GRBs). However, this behavior is also consistent with a particular spectral 
evolution expected from external shocks initiating an afterglow.

\section{Discussion}

The late, long-lag pulses of quiescent GRBs 960530 and 980125 have different 
temporal and spectral characteristics than those found earlier in the bursts. 
More notably, the properties are also different than those generally observed 
in GRBs. These properties indicate a separate pulse component which has been 
conveniently isolated from the standard emission in these two GRBs by the 
quiescent interval.

\subsection{Are the Long-Lag Components Internal or External Shocks?}

Despite the evidence linking GRB pulses to internal shocks, the long-lag 
pulses of GRBs 960530 and 980125 appear to be inconsistent with internal 
shocks. As can be seen in figure~\ref{fig1} and figure~\ref{fig2}, the 
long-lag pulses appearing late in the bursts are broader than the 
short-lag pulses appearing early in the bursts, as expected from 
external shock models \citep{fen96,sar97,fen99,frr99}. We thus suspect 
that this newly-identified emission component is an external shock 
signature.

Further evidence supporting the external shock hypothesis can be found 
from the relationship observed between the duration of the post-quiescent 
emission episode and the quiescent time \citep{rm01,rmr01}. These authors 
found a linear, one-to-one ratio which they interpreted as a metastable 
configuration in the outflow such that a significant energy fraction still 
stored in the fireball is available for liberation through additional 
internal shocks at a later time. Therefore a longer quiescent interval will
be followed by a longer post-quiescent emission episode. GRB 960530 and GRB 
980125 do not obey this relation. In GRB 960530, the ratio of the 
post-quiescent emission episode duration to the quiescent time is 
$37 {\rm s}/233 {\rm s} = 0.16$, while in GRB 980125 the ratio is 
$45 {\rm s}/161 {\rm s} = 0.28$. These values deviate considerably 
from those considered ``standard'' for quiescent GRBs. Because these 
values are not as expected for internal shocks, they could be examples 
of external shocks.

The newly-identified long-lag pulses have been found to occur towards the 
end of these GRBs; however, we note that the long-lag pulse in GRB 980125 
occurred prior to an additional concluding FRED pulse. (NOTE: both 
quiescent pulses have nearly the same peak flux.) The long-lag component 
may be present at even earlier times in other bursts. If so, then the 
newly-identified pulse type may be difficult to isolate because it is 
smooth, broad, and faint: it could overlap with narrower internal shock 
pulses to create complex temporal structures and complex spectral evolution. 
We note that many GRBs can be characterized by complex combinations of 
overlapping broad and narrow pulses.

\subsection{Impact on the Lag vs. Peak Luminosity Relation}

The lag vs. peak luminosity relation \citep{nor00} is
\begin{equation}
L_{53} \approx 1.3 \times \tau^{-1.14}
\end{equation}
where $\tau = ({\rm lag_{31}}/0.01 {\rm s})$, ${\rm lag}_{31}$ is the 
lag between the 100-300 keV channel and the 25-50 keV channel, and $L_{53}$ 
is the isotropic peak luminosity on the 256 ms timescale in units of $10^{53}$ 
erg s$^{-1}$. The relation was originally calibrated for a half-dozen GRBs of 
known redshift. However, the anomalously-underluminous long-lag GRB 980425 was 
not included in the calibration. 

The long-lag emission pulses of GRBs 960530 and 980125 provide a dilemma for 
the lag vs. peak luminosity relation, since that relation implicitly assumes 
that each GRB can be characterized by a single lag. However, very long 
post-quiescent lags do not put the relation at risk since there are normal 
bursts having even longer lags. According to figure 2 of \citet{nor02}, 
roughly 300 individual GRBs have ${\rm lag}_{21} > 0.43$ s, where 
${\rm lag}_{21}$ is the lag between the 50-100 keV channel and the 25-50 keV 
channel (${\rm lag}_{21} = 1.17 {\rm s}$ for GRB 960530 and 
${\rm lag}_{21} = 0.42 {\rm s}$ for GRB 980125). The long post-quiescent 
lags of GRBs 960530 and 980125 are common enough that they appear to 
represent a normal GRB behavior. 

It is possible that some bursts exhibit {\it only} the smooth, long lag 
(external shock) component. The variable, short lag (internal shock) 
component would most likely be undetected in these bursts. There is a 
distinct possibility, then, that \citet{nor02} has applied the lag vs.\ 
peak luminosity relation to bursts dominated by short-lag components as 
well as to potentially different bursts dominated by long-lag components. 
Quiescent GRBs 960530 and 980125 can provide a consistency check for GRBs 
that exhibit both behaviors. One would expect the post-quiescent to 
pre-quiescent luminosity ratios of these GRBs $as obtained from the lag
measurements$ to be equal to their peak flux ratios. 

For pre-quiescent episode $a$ and post-quiescent episode $b$, the lag vs.\ 
peak luminosity relation predicts a ratio of peak luminosities
\begin{equation}
\frac{L_b}{L_a} = (\frac{\tau_b}{\tau_a})^{-1.14}.
\end{equation}
For GRB 960530, $\tau_{\rm b}/\tau_{\rm a} = 8.7$, from which 
$L_{\rm b}/L_{\rm a} = 0.065$. However, the ratio of 256 ms 
peak fluxes gives $P_{\rm b}/P_{\rm a} = L_{\rm b}/L_{\rm a} = 0.46$. 
For GRB 980125, $\tau_{\rm b}/\tau_{\rm a} = 0.048$, which predicts 
$L_{\rm b}/L_{\rm a} = 0.024$. The ratio of 256 ms 
peak fluxes is $P_{\rm b}/P_{\rm a} = L_{\rm b}/L_{\rm a} = 0.15$.

\begin{figure}
%\plotone{laglum_sn_arrows.ps}
\plotone{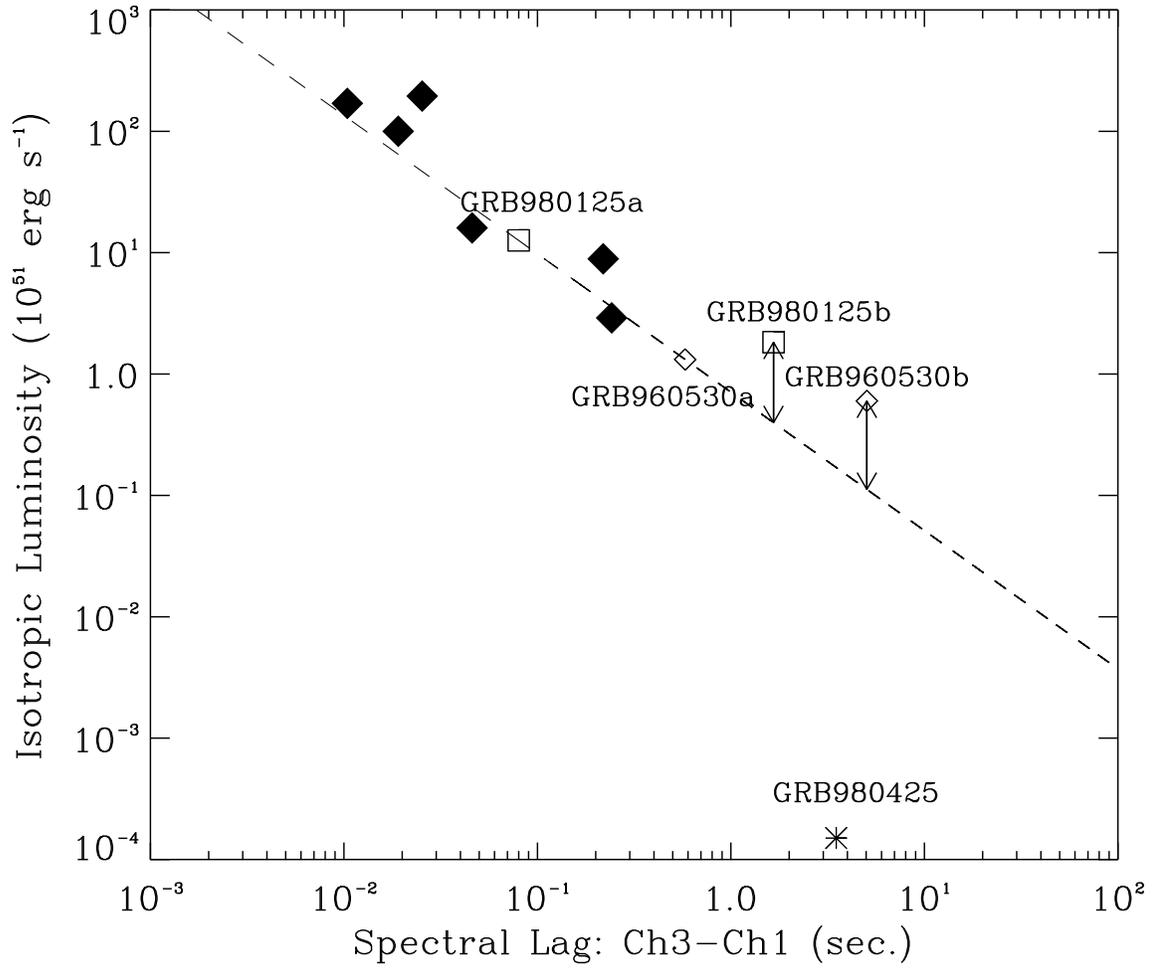}
\caption{The lag vs. peak luminosity relation applied to GRBs 960530 and 980125).\label{fig4}}
\end{figure}

The peak luminosities of the short-lag (a) and long-lag (b) components 
are plotted in figure~\ref{fig4} together with the six short-lag GRBs 
(filled diamonds) used to calibrate the lag vs.\ peak luminosity relation 
\citep{nor00}. The luminosities of the long-lag components are predicted 
using both the lag vs.\ peak luminosity relation and peak flux ratios; 
the values predicted by the lag vs. peak luminosity relation lie on the 
theoretical (dashed) line while those predicted from peak flux ratios are 
shifted vertically from this line. The luminosities calculated from peak 
fluxes are larger than those calculated from the lag vs. peak luminosity 
relation, suggesting that the luminosities of these long-lag components 
are overluminous relative to the lag vs. peak luminosity relation.

This is difficult to reconcile with long-lag GRB 980425, which is 
underluminous relative to the lag vs. peak luminosity relation (see 
figure~\ref{fig4}) rather than overluminous. Even though GRB 980425 
and the post-quiescent emission of GRBs 960530 and 980125 all have 
long lags of similar durations, it appears that these lags are not 
good predictors of peak luminosity. The long-lag GRBs do not all 
obey the same lag vs. peak luminosity relation as short-lag GRBs, 
and it is possible that they do not obey a lag vs.\ peak luminosity 
relation at all.

\subsection{How Common is the Long-Lag Emission Component?}

The long-lag emission component is likely common in many GRBs, even though 
it has, thus far, only been expicitly identified in only two quiescent GRBs. 
The lag calculations of \cite{nor00} use an apodization technique to remove 
low-intensity emission from the lag calculation because a long-lag signature 
hidden in the low-intensity emission is capable of distorting the lag 
measurement (e.g. GRB 990123). The authors thus implicitly recognize the 
presence of a long-lag component arising from low-intensity emission. 
In addition to this signature, we have seen evidence of additional 
long-lag emission in the decaying tail of GRB 991216 (BATSE trigger 7906). 
As indicated by \cite{nor02}, it is also possible that many GRBs consist 
of a long-lag component only.

\subsection{Impact on Theoretical Models}

The two different pulse types observed here presumably arise from the same 
collimated relativistic outflow. The narrow pulses originate from colliding 
shells in the high-$\Gamma$, tightly-collimated jet center (e.g. 
\cite{mrw98,kum00}), where $\Gamma$ is the bulk Lorentz factor. The external 
shocks initiating the onset of afterglow are produced by a low-$\Gamma$, 
slowly-expanding shock found at the interface between the jet and the external 
medium (i.e. the external shock). The opening angle of the jet should increase 
as deceleration sets in (e.g. \citet{rrm99}). The detailed structure of GRB 
relativistic jets is still unclear, but determination of this structure is 
currently being performed using numerical hydrodynamic codes (e.g. 
\citep{zha03,alo03}). In fact, hydrodynamic simulations indicate that the 
beaming angle of the high-$\Gamma$ component is limited to $\sim 5^\circ$, 
whereas the low-$\Gamma$ component can have a beaming angle exceeding 
$15^\circ$ \cite{zha03}. The angular expansion of the external shock can be 
aided by the presence of a ``cocoon jet'', which is a secondary jet structure 
that forms when energy is deposited in a stellar cavity prior to breakout 
\cite{rcr02}.

In the standard internal/external shock model, deceleration of the 
relativistic ejecta sets in at some deceleration distance 
$R_{dec} \sim \Gamma^{-2/(3-s)}$ in the surrounding environment with 
density $n(r) \sim r^{-s}$, resulting in a temporal delay between the 
prompt internal shock $\gamma$-ray emission and the afterglow \citep{rcr02}. 
Thus, for a given density profile, large $\Gamma$ yields a small $R_{dec}$. 
A temporal overlap between the internal and external shock emission would 
be expected in these cases, producing complex $\gamma$-ray light curves. 
The optical flash that occurred some $\sim 40$ s into the prompt $\gamma$-ray 
emission of GRB 990123 has been interpreted as deceleration during the burst 
due to the reverse external shock \citep{sap99,mr99,srr02}. Interestingly, 
the optical flash is temporally coincident with the long-lag component of 
GRB 990123. \citet{sap99} note that the x-ray emission peaked some 
$\sim 60$ s into the burst. This overlaps in time with the long-lag 
emission, suggesting that the long-lag component may represent the onset 
of the afterglow, e.g. the long-lived forward shock emission. On the other 
hand, the low-intensity long-lag emission exhibits some short timescale 
variability. \citet{frb99} demonstrated that the widths of these low-intensity 
pulses do not evolve with time, suggesting no deceleration and an internal 
shock origin. However, this does not preclude the possibility of narrow 
late-time pulses overlapping a broad long-lag emission component. We note 
that these behaviors are quite unlike the behaviors observed in that latter 
parts of GRBs 950530 and 980125.

As noted before, the BATSE database contains many long-lag GRBs. These 
often have characteristics of FREDs, and many are initiated by what appear 
to be narrow, short-lag pulses. It is tempting to interpret these as GRBs 
produced by jets with small $R_{dec}$; few internal shocks are observed, 
and the onset of afterglow is thus the defining characteristic of their 
prompt emission. However, this hypothesis is complicated by the menagerie 
of other phenomena such as X-ray rich GRBs (XRRs), X-Ray Flashes(XRFs), 
and orphan afterglows, that are also thought to result from off-axis GRB 
viewing (e.g. \citet{mrw98,zha04}). This is because XRRs and XRFs are 
spectrally soft, but otherwise contain many pulses; some of these pulses 
are narrow and some are broad (R. Vanderspek, private communication). Our 
explanation of the long-lag bursts would imply that XRRs and XRFs are jets 
with small viewing angles (to explain the variable temporal structure) seen 
at large redshift, rather than being GRB jets viewed off-axis. However, 
\citet{blo03} have shown that two recent XRFs originated at redshifts 
less than $z = 3.5$.

\section{Conclusions}

At least two quiescent GRBs exhibit multiple lags. The short-lag component 
is consistent with internal shocks while an external shock explanation is 
preferred for the smooth long-lag component. The long-lag component is broad 
and faint, and is possibly present but unrecognizable in many bursts. The 
apodization of \citep{nor00} implicitly recognizes its existence in order 
to remove it. It is possible that many or all of the long-lag bursts in the 
sample of \citet{nor02} exhibit only the smooth and broad long-lag component.

Because the short and long lags found in these quiescent GRBs are so different,
we have assumed that there are two distinct GRB pulse types. This hypothesis 
needs to be verified. The best way to approach this problem is via statistical 
and/or data mining studies of GRB pulse characteristics. It is important to 
determine the existence of one continuous pulse distribution or poossibly 
distinctly different distributions. Detailed analyses of BATSE and $Swift$ 
GRB data are planned to corroborate this work.

The lag vs. peak luminosity relation apparently requires re-calibration. It 
is not clear that long-lag pulses obey the same relation as short-lag pulses, 
or if they even obey a lag vs. peak luminosity relation. Since there is a 
correlation between lag and variability, the variability vs. peak luminosity 
relation \citep{rei01} might require similar re-calibration.

The simplest explanation for the large number of long-lag GRBs found in the 
BATSE database is that they are GRBs viewed from off the jet axis. The 
tightly-beamed, variable, short-lag internal shocks would not be seen from 
large viewing angles, whereas the more widely-beamed, smooth, long-lag 
external shocks would be. However, it is difficult to reconcile this 
explanation at the present time with some of the other similar phenomena 
associated with GRBs (e.g. x-ray rich GRBs, Gamma-Ray Flashes, and Orphan 
Afterglows).




%% Included in this acknowledgments section are examples of the

%% AASTeX hypertext markup commands. Use \url without the optional [HREF]

%% argument when you want to print the url directly in the text. Otherwise,

%% use either \url or \anchor, with the HREF as the first argument and the

%% text to be printed in the second.

\section{Acknowledgments}

We are grateful to the anonymous referee for a careful reading of this 
manuscript, and for suggestions that strengthened the theoretical 
discussion. We acknowledge R. D. Preece for calculating burst locations, 
and we are extremely grateful to T. Piran, E. E. Fenimore, E. Ramirez-Ruiz, 
J. Norris, J. Bonnell, and R. Vanderspek for valuable discussions. This 
research was sponsored by NASA grant NRA-98-OSS-03 and NSF grant AST00-98499.






%%

















\clearpage


\clearpage

\clearpage

\clearpage

%% If you are not including electonic art with your submission, you may

%% mark up your captions using the \figcaption command. See the 


%%

%% No more than seven \figcaption commands are allowed per page, 

%% so if you have more than seven captions, insert a \clearpage 


%% Tables should be submitted one per page, so put a \clearpage before


\clearpage

\begin{deluxetable}{lccccc}
\tabletypesize{\scriptsize}
\tablecaption{Properties of Quiescent GRB 960530 (BATSE trigger 5478). \label{tbl-1}}
\tablewidth{0pt}
\tablehead{
\colhead{Time Period} & \colhead{morphology}   & \colhead{ILF $\alpha$}   &
\colhead{Ch 21 lag (sec)} &
\colhead{Ch 31 lag (sec)} & \colhead{Ch 32 lag (sec)}
}
\startdata
before 200 sec & spiky FRED & $0.9 \pm 0.4$ & $0.1 \pm 0.11$ & $0.58 \pm 0.12$ & $0.58 \pm 0.04$ \\
after 200 sec & smooth gradual & $-7.5 \pm 1.2$ & $1.17 \pm 0.26$ & $5.04 \pm 0.06$ & $2.77 \pm 0.06$ \\
\enddata
\end{deluxetable}

\clearpage

\begin{deluxetable}{lccccc}
\tabletypesize{\scriptsize}
\tablecaption{Properties of Quiescent GRB 980125 (BATSE Trigger 6581). \label{tbl-2}}
\tablewidth{0pt}
\tablehead{
\colhead{Time Period} & \colhead{morphology}   & \colhead{ILF $\alpha$}   &
\colhead{Ch 21 lag (sec)} &
\colhead{Ch 31 lag (sec)} & \colhead{Ch 32 lag (sec)}
}
\startdata
before 150 sec & complex + FRED & $-1.3 \pm 0.5$ & $0.04 \pm 0.02$ & $0.08 \pm 0.01$ & $0.10 \pm 0.01$ \\ 
150 to 250 sec & smooth gradual & $-3.1 \pm 0.5$ & $0.43 \pm 0.32$ & $1.66 \pm 0.19$ & $0.90 \pm 0.40$ \\
after 250 sec & smooth FRED & $-3.9 \pm 0.4$ & $0.07 \pm 0.03$ & $0.13 \pm 0.08$ & $0.05 \pm 0.08$ \\
\enddata
\end{deluxetable}

%% The following command ends your manuscript. LaTeX will ignore any text



\begin{thebibliography}{}

\bibitem[Aloy et al.(2003)]{alo03} Aloy, M.-A., Mart\'{i}, J.-M., G\'{o}mez, J.-L., Agudo, I., M\"{u}ller, E. \& Iban\'{e}z, J.-M.\ 2003, \apj, 585, L109
\bibitem[Band(1997)]{ban97} Band, D.~L.\ 1997, \apj, 486, 928
\bibitem[Bloom et al.(2003)]{blo03} Bloom, J.~S., Fox, D., 
van Dokkum, P.~G., Kulkarni, S.~R., Berger, E., Djorgovski, S.~G., \& 
Frail, D.~A.\ 2003, \apj, 599, 957
\bibitem[Burenin et al.(1999)]{bur99} Burenin, R.~A.~et al.\ 
1999, \aap, 344, L53
\bibitem[Connaughton(2002)]{con02} Connaughton, V.\ 2002, 
\apj, 567, 1028 
\bibitem[Costa(2000)]{cos00} Costa, E.\ 2000, AIP 
Conf.~Proc.~526: Gamma-ray Bursts, 5th Huntsville Symposium, 365
\bibitem[Daigne \& Mochkovitch(1998)]{dm98} Daigne, F.~\& 
Mochkovitch, R.\ 1998, \mnras, 296, 275
\bibitem[Fenimore, Madras, ~\& Nayakshin(1996)]{fen96} 
Fenimore, E.~E., Madras, C.~D., \& Nayakshin, S.\ 1996, \apj, 473, 998 
\bibitem[Fenimore et al.(1999)]{fen99} Fenimore, E.~E., 
Cooper, C., Ramirez-Ruiz, E., Sumner, M.~C., Yoshida, A., ~\& Namiki, M.\ 
1999, \apj, 512, 683 
\bibitem[Fenimore \& Ramirez-Ruiz(1999)]{frr99} Fenimore, 
E.~E.~\& Ramirez-Ruiz, E.\ 1999, ASP Conf.~Ser.~190: Gamma-Ray Bursts: The 
First Three Minutes, 67 
\bibitem[Fenimore, Ramirez-Ruiz, \& Wu(1999)]{frb99} Fenimore, E.~E., Ramirez-Ruiz, E., \& Wu, B.\ 1999, \apjl, 518, L73
\bibitem[Frail et al.(2001)]{fra01} Frail, D.~A.~et al.\ 
2001, \apjl, 562, L55
\bibitem[Freismuth, Giblin, ~\& Hakkila(2003)]{fre03} 
Freismuth, T.~M., Giblin, T., ~\& Hakkila, J.\ 2003, American Astronomical 
Society Meeting, 202
\bibitem[Giblin et al.(1999)]{gib99} Giblin, T.~W., van 
Paradijs, J., Kouveliotou, C., Connaughton, V., Wijers, R.~A.~M.~J., 
Briggs, M.~S., Preece, R.~D., ~\& Fishman, G.~J.\ 1999, \apjl, 524, L47 \bibitem[Giblin(2000)]{gib00} Giblin, T.~W.\ 2000, Ph.D.~Thesis
\bibitem[Giblin et al.(2002)]{gib02} Giblin, T.~W., 
Connaughton, V., van Paradijs, J., Preece, R.~D., Briggs, M.~S., 
Kouveliotou, C., Wijers, R.~A.~M.~J., ~\& Fishman, G.~J.\ 2002, \apj, 570, 
573 
\bibitem[Hakkila et al.(2004)]{hak04} Hakkila, J., Giblin, T.~W., Young, K.~C., Fuller, S.~P., Stallworth, A.~D., \& Sprague, A.~P., 2004, in proceedings of the 2003 Santa Fe Gamma-Ray Burst Conference (ed. E.~E. Fenimore), submitted.
\bibitem[Horack \& Hakkila(1997)]{hor97} Horack, J.~M.~\& Hakkila, J.\ 1997, \apj, 479, 371
\bibitem[Kobayashi, Piran, \& Sari(1997)]{kss97} Kobayashi, 
S., Piran, T., ~\& Sari, R.\ 1997, \apj, 490, 92 
\bibitem[Kumar \& Piran(2000)]{kum00} Kumar, P.~\& Piran, T.\ 
2000, \apj, 535, 152
\bibitem[Meegan et al.(2003)]{mee03} Meegan, C.~A. et al. 2003, Current BATSE Gamma-Ray Burst Catalog. Available at gammaray.msfc.nasa.gov/batse/grb/catalog
\bibitem[Nakar \& Piran(2002)]{np02} Nakar, E.~\& Piran, T.\ 
2002, \apjl, 572, L139
\bibitem[M\'{e}sz\'{a}ros, Rees, \& Wijers(1998)]{mrw98} M\'{e}sz\'{a}ros, P., Rees, M.~J. \& Wijers,~R.~A.~M.~J.\ 1998, \apj, 499, 301
\bibitem[M\'{e}sz\'{a}ros, \& Rees(1999)]{mr99} M\'{e}sz\'{a}ros, P., \& Rees, M.~J.\ 1999, \mnras, 306, L39
\bibitem[Norris(2002)]{nor02} Norris, J.~P.\ 2002, \apj, 579, 386 
\bibitem[Norris et al.(1996)]{nor96} Norris, J.~P., Nemiroff, 
R.~J., Bonnell, J.~T., Scargle, J.~D., Kouveliotou, C., Paciesas, W.~S., 
Meegan, C.~A., \& Fishman, G.~J.\ 1996, \apj, 459, 393 
\bibitem[Norris, Marani, ~\& Bonnell(2000)]{nor00} Norris, 
J.~P., Marani, G.~F., ~\& Bonnell, J.~T.\ 2000, \apj, 534, 248
\bibitem[Norris, Scargle, \& Bonnell(2001)]{nor01} Norris, 
J.~P., Scargle, J.~D., ~\& Bonnell, J.~T.\ 2001, AIP Conf.~Proc.~587: Gamma 
2001: Gamma-Ray Astrophysics, 176
\bibitem[Paciesas et al.(1999)]{pac99} Paciesas, W.~S.~et al.\ 1999, \apjs, 122, 465
\bibitem[Panaitescu \& Kumar(2002)]{pan02} Panaitescu, A.~\& 
Kumar, P.\ 2002, \apj, 571, 779
\bibitem[Ramirez-Ruiz, Celotti, \& Rees(2002)]{rcr02} Ramirez-Ruiz, E., Celotti, A., ~\& Rees, M.~J.\ 2002, \mnras, 337, 1349
\bibitem[Ramirez-Ruiz \& Fenimore(1999)]{rrm99} Ramirez-Ruiz, 
E.~\& Fenimore, E.~E.\ 2000, \aaps, 138, 521
\bibitem[Ramirez-Ruiz \& Fenimore(2000)]{rrf00} Ramirez-Ruiz, 
E.~\& Fenimore, E.~E.\ 2000, \apj, 539, 712
\bibitem[Ramirez-Ruiz \& Merloni(2001)]{rm01} Ramirez-Ruiz, 
E.~\& Merloni, A.\ 2001, \mnras, 320, L25
\bibitem[Ramirez-Ruiz, Merloni, ~\& Rees(2001)]{rmr01} 
Ramirez-Ruiz, E., Merloni, A., ~\& Rees, M.~J.\ 2001, \mnras, 324, 1147
\bibitem[Reichart et al.(2001)]{rei01} Reichart, D.~E., Lamb, 
D.~Q., Fenimore, E.~E., Ramirez-Ruiz, E., Cline, T.~L., ~\& Hurley, K.\ 
2001, \apj, 552, 57
\bibitem[Rees \& M\'{e}sz\'{a}ros(1994)]{ree94} Rees, M.~J.~\& 
M\'{e}sz\'{a}ros, P.\ 1994, \apjl, 430, L93
\bibitem[Sari(1999)]{sar99} Sari, R.\ 1999, \apjl, 524, L43
\bibitem[Sari \& Piran(1997a)]{sar97a} Sari, R.~\& Piran, T.\ 
1997a, \mnras, 287, 110
\bibitem[Sari \& Piran(1997b)]{sar97} Sari, R.~\& Piran, T.\ 
1997b, \apj, 485, 270
\bibitem[Sari \& Piran(1999)]{sap99} Sari, R.~\& Piran, T.\ 
1999, \apj, 517, L109
\bibitem[Soderberg \& Ramirez-Ruiz(2002)]{srr02} Soderberg, A.~M., \& Ramirez-Ruiz, E.~\ 2002, \mnras, 330, L24
\bibitem[Stern, Poutanen, \& Svensson(1999)]{ste99} Stern, 
B., Poutanen, J., \& Svensson, R.\ 1999, \apj, 510, 312
\bibitem[Wu \& Fenimore(2000)]{wu00} Wu, B.~\& Fenimore, E.\ 
2000, \apjl, 535, L29
\bibitem[Zhang et al.(2004)] {zha04} Zhang, B., Dai, X., Lloyd-Ronning, N. ~M., ~\& M\'{e}sz\'{a}ros, P.\ 2004, \apj, 601, L119
\bibitem[Zhang, Woosley, \& MacFadyen(2003)]{zha03} Zhang, 
W., Woosley, S.~E., \& MacFadyen, A.~I.\ 2003, \apj, 586, 356
\end{thebibliography}
\end{document}